\documentclass[a4paper,11pt]{article}
\usepackage[margin=2.5cm]{geometry}
\usepackage{graphicx}
\usepackage{epsfig}
\usepackage{amsmath}
\usepackage{amssymb}
\usepackage{cite}
\usepackage{float}
\usepackage[title]{appendix}
\usepackage[font={footnotesize,it}]{caption}
\begin{document}

%%%%%%%%%%%%%%%%%%%%%%%%%%%%%%%%%%%%%%%%%%%%%%%%%%%%%%%%%%%%%%%%%%%%%%%%%%%%%%%%%%%%%%%%
\begin{titlepage}
\title{Thermodynamic Geometry and Phase Transitions of AdS Braneworld Black Holes}
\author{}
\date{% authors are dated
Pankaj Chaturvedi and Gautam Sengupta
\thanks{\noindent E-mail:~ cpankaj@iitk.ac.in , sengupta @iitk.ac.in}
\vskip0.4cm
{\sl Department of Physics, \\
Indian Institute of Technology,\\
Kanpur 208016, \\
India}}
\maketitle
\abstract{
\noindent The thermodynamics and phase transitions of charged RN-AdS and rotating Kerr-AdS black holes in a generalized Randall-Sundrum braneworld are investigated in the framework of thermodynamic geometry.  A detailed analysis of the thermodynamics, stability and phase structures in the canonical and the grand canonical ensembles for these AdS braneworld black holes are described. The thermodynamic curvatures for both these AdS braneworld black holes are computed and studied as a function of the thermodynamic variables. Through this analysis we illustrate an interesting dependence of the phase structures on the braneworld parameter for these black holes.
}

\end{titlepage}
%%%%%%%%%%%%%%%%%%%%%%%%%%%%%%%%%%%%%%%%%%%%%%%%%%%%%%%%%%%%%%%%%%%%%%%%%%%%%%%%%%%%%%%%

\section{Introduction}
In recent years black hole thermodynamics has emerged as a theoretical laboratory for testing the predictions of candidate theories of quantum gravity. In this context this field has witnessed significant focus of research in the recent past that is reviewed in \cite{Hawking:1974sw, Bekenstein:1973ur, Bardeen:1973gs, Wald:1999vt, Page:2004xp, Ross:2005sc, Hawking:1982dh,Townsend:1997ku,Dabholkar:2012zz}.  Black holes have been established as thermodynamic objects with a Bekenstein-Hawking entropy and a Hawking temperature. Although a full microscopic statistical description for black hole is yet incomplete, thermodynamic analysis has led to a rich structure of phase transitions and critical phenomena. In particular, inspired by the gauge-gravity correspondence the central focus of these investigations have involved the study of the thermodynamics of asymptotically Anti deSitter (AdS) black holes. The AdS black holes are specifically suited for a thermodynamic analysis as they are stable in the canonical ensemble over large ranges of their thermodynamic parameters unlike their asymptotically flat counterparts. Notice however that the thermodynamics and phase structures of black holes, unlike conventional thermodynamic systems exhibit a dependence on the ensemble considered. This is characteristic of all self gravitating systems which cannot be partitioned into non interacting subsystems. Despite this important difference the thermodynamics and phase structures of black holes remarkably exhibit interesting similarities to conventional thermodynamic systems. It was shown in \cite{Hawking:1982dh} that there was a Hawking-Page phase transition for AdS-Schwarzschild black holes between a thermal AdS space time and a AdS black hole at some temperature. Whereas the specific heat has a divergence where it changes sign displaying a turning point behaviour which is termed as Davies points. The charged RN-AdS black holes show an even more interesting phase structure with a Hawking-Page transition for the grand canonical ensemble and a first order liquid gas transition ending in a critical point for the canonical ensemble \cite {Chamblin:1999tk, Chamblin:1999hg}. Similar variety of phase structures are also exhibited by the Kerr-AdS and the Kerr-Newman-AdS black holes in various ensembles as described in \cite{Caldarelli:1999xj}.

Over the last few decades it has also become increasingly clear that higher dimensional space-time has assumed considerable significance for low energy effective field theories arising from fundamental theories of gravity. The extra space time dimensions in these models are assumed to be compactified to small scales to provide the effective low energy descriptions. This relegates the quantum effects of gravity to very high energy scales like the Planck energy. However one of the interesting constructions in the last decade has been the advent of the brane world scenarios characterized by large compactification scales which could then admit low scale effects of quantum gravity \cite {ArkaniHamed:1998rs,ArkaniHamed:1998nn,Antoniadis:1998ig,Rubakov:1983bb,Rubakov:1983bz,Visser:1985qm,Maartens:2010ar,Dick:2001sc}. This naturally had very interesting associated phenomenology and was the focus of intense interest. Such braneworld constructions involved the restriction of the gauge interactions to a lower dimensional hypersurface (brane) embedded in a higher dimensional bulk space time with the electroweak scale as the fundamental scale.  Thus the four dimensional Planck scale was a derived scale in this scenario leading to a possible resolution of the gauge hierarchy problem. The most well known of such brane world construction were the Randall-Sundrum (RS) models \cite {Randall:1999ee,Randall:1999vf} which involved a three brane embedded in a bulk space time which was a slice of a five dimensional Anti deSitter (AdS) space with a warped compactification. Generic four dimensional gravitational configurations on the brane could then be described from a higher dimensional perspective \cite{Garriga:1999yh,Shiromizu:1999wj,Giddings:2000mu}. The investigation of  black hole configurations in four dimensions from a higher dimensional bulk perspective was a significant aspect of such studies in braneworld gravity \cite {Cavaglia:2002si,Kanti:2004nr,Giannakis:2000zx,Emparan:1999wa,Kanti:2001cj,Casadio:2002uv,Emparan:2002px,Chamblin:1999by,Gregory:2000gf,Dadhich:2000am,Emparan:1999fd,Modgil:2001hm,Cardoso:2004zz,Aliev:2005bi,Chamblin:2000ra,Shiromizu:2000pg,Sengupta:2002fr,Sahay:2007rm}. However recently the authors in \cite{PhysRevLett.117.011102} have made a first numerical study of the full dynamics of a braneworld scenario. In particular they have established black hole formation in Randall-Sundrum II braneworld model, where braneworld black hole solutions are obtained from dynamical evolutions, these black holes have finite extension  into the bulk and they are consistent with those previously computed from a static vacuum ansatz.

In the original Randall-Sundrum braneworld construction, the 3-brane on which the gauge interactions were restricted was considered to have a zero cosmological constant. While this is consistent with the resolution of the gauge hierarchy problem which these models addressed, it was incompatible with the obersvational evidence of the cosmological constant being small but non zero and positive. Thus there was a necessity to generalize the RS construction to obtain a non zero cosmological constant on the 3-brane. Such a generlaization was attempted in \cite {Das:2007qn} and it could be shown that both postive and negative cosmologican constant 3-branes were admissible in the RS framework leading to both De Sitter (dS) as well as Anti deSitter (AdS)braneworlds. Interestingly while a dS solution led to negative brane tension the corresponding AdS braneworld solutions involved both positive and negative tension branes. Remarkably for the AdS solution the resolution for the gauge hierachy problem induces an upper bound on the brane cosmological constant. It was a naturally interesting proposition to study black holes in such AdS braneworlds. In particular it was an interesting issue to investigate the thermodynamics of such AdS braneworld black holes in such a generalized RS scenario. In \cite {Koley:2009gx} the thermodynamics and phase transitions of such AdS-Schwarzschild (SAdS) and Reissner-Nordstrom-AdS (RNAdS) braneworld black holes were investigated. It was shown that the thermodynamic variables exhibited interesting dependence on the braneworld parameters namely the bulk and the brane cosmological constants. Study of the specific heats also showed the dependence of the Hawking Page phase transition on the braneworld parametrs.

In a different context a geometrical approach to thermodynamics, phase transitions and critical phenomena has been developed over the last few decades. Following from the early work of Weinhold \cite {Weinhold:1975sc, Weinhold:1975hp}, a Riemannian geometrical structure with a Euclidean signature was established in the equilibrium thermodynamic state space of a system by Ruppeiner \cite {Ruppeiner:1995zz, Ruppeiner:1983zz}. The positive definite invariant line element of this geometry  was shown to be related in a Gaussian approximation to the probability measure of the fluctuations in the thermodynamic variables in a specific representation. Most significantly the scalar curvature computed from this geometry encoded the interactions in  the underlying microscopic statistical system. For a non interacting system like the ideal gas the thermodynamic curvature was computed to be zero describing a flat thermodynamic geometry. From simple scaling and hyperscaling arguments it could be shown that the thermodynamic scalar curvature scaled like the correlation volume of the system. Hence at a critical point where the correlation length diverges the thermodynamic scalar curvature develops a singularity and may be used to characterize the second order phase transition. Application of this geometric framework to conventional thermodynamic systems showed a remarkable correspondence with their thermodynamic behaviours and phase structures. The connection between the macroscopic thermodynamic description and the microscopic interactions afforded by this framework makes makes it extremely suitable to describe the thermodynamics of black holes where the microscopic description is unavailable. In fact application of this geometric framework to describe the thermodynamics of both extremal and non extremal black holes have led to interesting new physical and mathematical insights \cite {Gibbons:1996af, Ferrara:1997tw, Ruppeiner:2007hr, Ruppeiner:2008kd,Aman:2003ug, Cai:1998ep, Sarkar:2008ji, Sarkar:2006tg,Banerjee:2010bx, Banerjee:2010da, Akbar:2011qw}. This geometric framework which could describe second order phase transitions through the divergence of the thermodynamic scalar curvature was subsequently generalized to describe first order phase transitions and supercritical phenomena by one of the authors in a series of collaborations \cite {Sahay:2010tx, Sahay:2010wi, Sahay:2010yq, Ruppeiner:2011gm}. The generalization involved the study of the branching behaviour of the thermodynamic scalar curvature as a function of thermodynamic variables like the temperature and led to the R-Crossing formula as an alternative to the usual Maxwell-van der Waals criteria of equal free energies for coexisting phases. The thermodynamic curvature also served as a first theoretical description for the Widom line in the supercritical regime describing the locus of the maxima of the correlation length beyond the critical point. Thus this geometric framework provided a unified geometrical characterization of all subcritical, critical and supercritical phenomena.

Given the background of such interesting insights obtained for the thermodynamics of AdS black holes using the framework of thermodynamic geometries, it is a natural question to study the thermodynamics and phase transitions of the AdS braneworld black holes obtained in a generalized RS scenario mentioned earlier. In particular this geometric framework may lead to interesting new insights into the thermodynamics and phase transitions of such AdS braneworld black holes. In this article we investigate in details the thermodynamics and phase transitions of both the charged RN-AdS and the rotating Kerr-AdS braneworld black holes in the framework of thermodynamic geometries. Given the ensemble dependence of the phase behaviours of black holes we implement a careful analysis for both the canonical and the grand canonical ensembles. The thermodynamic scalar curvature for these black holes are computed and studied as a function of the temperature to investigate their phase structures. We observe the expected divergence and turning point behaviours of the thermodynamic scalar curvature for the Hawking Page and the Davies transitions in accordance with the observations in \cite {Sahay:2010tx}. Interestingly the thermodynamic scalar curvature as a function of temperature exhibits a dependence on the braneworld parameters.

%%%%%%%%%%%%%%%%%%%%%%%%%%%%%%%%%%%%%%%%%%%%%%%%%%%%%%%%%%%%%%%%%%%%%%%%%%%%%%%%%%%%%%%%
\section{Randall Sundurum Model For Braneworld  With  Non-Zero Cosmological Constant}
The usual Randall-Sundrum (RS) construction with negative cosmological constant involves two 3-branes embedded in a slice of an $AdS_{5}$ bulk space time separated by a distance in the extra fifth direction. The extra spatial dimension is compactified on a $S_{1}/Z_{2}$ orbifold with reflection symmetric boundary conditions at the location of the three brane. One of the three branes correspond to the visible universe and the second is a regulator brane. In a subsequent variant the second 3-brane may be relegated to infinity thus leading to a model with a single 3-brane and an infinite extra fifth dimension.  The metric for Randall-Sundrum model with such a warped compactification may be expressed as,
\begin{equation}
ds^{2}=e^{-2kr_{c}|y|}\eta_{\mu\nu}dx^{\mu}dx^{\nu}+r_{c}^{2}dy^{2},\label{metric1}
\end{equation}
here $k$ is of the oder of Planck scale $\left(M_{p}\right)$ and $r_{c}$ is the radius for extra fifth spatial dimension. The three branes reside at the fixed points $y=\pi$ (visible) and $y=0$ (hidden) of the orbifold $S_{1}/Z_{2}$ geometry. The cosmological constant on the 3-brane is considered to be zero in the standard RS framework. This may be generalized through the use of a more general warp factor which leads to both positive and negative cosmological constants on the 3-brane. Naturally such a  generalization admits of non Ricci flat gravitational configurations on the 3-brane. Thus it is possible to have both de-Sitter (dS) and Anti de-Sitter (AdS) black hole solutions on the 3-brane. The generalized form for warped metric may be given  as \cite{Das:2007qn},
\begin{equation}
ds^{2}=e^{-2A(y)}g_{\mu\nu}dx^{\mu}dx^{\nu}+dy^{2}.\label{metric2}
\end{equation}

The 3-branes for this construction are as usual located at the fixed points $y=0$ and $y=\pi$ with the generalized wrap factor $e^{-2A(y)}$ evaluated as,
\begin{equation}
e^{-2A(y)}=\omega^{2}\cosh^{2}(\ln\frac{\omega}{c_{1}}+ky),\label{warpfac}
\end{equation}
where, $\omega^{2}=-\frac{\Theta}{3k^{2}}$ also, $c_{1}=1+\sqrt{1-\omega^{2}}$. Here $\Theta$  is the cosmological constant of the 3-branes and the sign of $\omega^{2}$ determines the asymptotic AdS or dS nature of 3-brane. The remaining factor $k$ is related to the cosmological constant $(\Lambda)$ of the bulk space time with $k\sim\sqrt{-\frac{\Lambda}{12M^{3}}}$ and $ M$ as the value for the mass in the bulk space time. The metric $g_{\mu\nu}$ is the metric for the gravitational configuration on the 3-brane \footnote{A detailed explanation on how to obtain the braneworld black hole solutions is given in the appendix A}.

%%%%%%%%%%%%%%%%%%%%%%%%%%%%%%%%%%%%%%%%%%%%%%%%%%%%%%%%%%%%%%%%%%%%%%%%%%%%%%%%%%%%%%%%
\section{Thermodynamic Geometry of Charged-AdS BraneWorld Black Holes}
In this section we consider charged black holes on the 3-brane in such generalized AdS braneworld models with a negative cosmological constant. The metric for such a of charged AdS brane world black hole may be expressed as \cite {Koley:2009gx},
\begin{equation}
ds^{2}=e^{-2A(y)}[-f(r)dt^{2}+\frac{dr^{2}}{f(r)}+r^{2}d\Omega^{2}]+dy^{2},\label{rnmetric}
\end{equation}
where,\begin{equation}
f(r)=1-\frac{2M}{M_p^{2}}\frac{1}{r}+\frac{Q^{2}}{M_p^{2}}\frac{1}{r^{2}}+k^{2}\omega^{2}r^{2}.\label{rnlapse}
\end{equation}

This further leads to a 3-brane with a constant negative curvature which admits a charged AdS black hole solution on the brane with a mass $M$ and charge $Q$. The event horizon for the charged AdS black hole on the 3-brane is fixed from the solution of the equation $f(r)=0$.  The preceding equation leads to four roots out of which only two roots may be considered as physical with $r_{+}$ (corresponding to the outer horizon) and $r_{-}$ (corresponding to the inner horizon). Out of the two radii, the outer horizon radii $r_{+} = r_{H}$ is usually considered for calculating the entropy and the temperature of  the charged AdS black hole. Now we modify the equation (\ref{rnlapse}) by putting, $q=\frac{Q}{Mp}, m=\frac{M}{Mp^{2}}$ and set the scale $ k=1$  which leads to the following expression,
\begin{equation}
f=1-\frac{2m}{r}+\frac{q^{2}}{r^{2}}+r^{2}\omega^{2}.\label{rnlapse1}
\end{equation}
here the Maxwell field equations are satisfied if one considers the ansatz $A=\phi(r) dt$ with $\phi(r)=-q/r$ for the gauge field on the 3- brane. Using the above equation we can determine the temperature $\left(T\right)$, electric  potential $\left(\phi\right)$, specific heat at constant charge $q$ $\left(C_{q}\right)$ and specific heat at constant electric potential $\phi$ as $\left(C_{\phi}\right)$. The temperature $T$ as a function of charge $(q)$ and entropy $(S)$ is given as,
\begin{equation}
T=\frac{-\pi ^2 q^2+\pi  S+3 S^2 \omega ^2}{4 \pi ^{3/2} S^{3/2}}. \label{rntemp}
\end{equation}
where the electric potential $\phi$ is a function of the scaled charge $q$. Correspondingly the entropy $S$ may be expressed as,
\begin{equation}
\phi =  \frac{\sqrt{\pi } q}{\sqrt{S}}. \label{rnphi}
\end{equation}

The value of the chemical potential $\phi$ considered here is the potential measured at infinity with respect to the horizon. The expressions for the specific heat capacity at constant charge denoted by $C_{q}$ and the heat capacity at constant potential   $C_{\phi}$ as a function of $q$ and $S$ are given as,
\begin{eqnarray}
C_{q}=\frac{2 S \left(3 S^2 \omega ^2+\pi  S-\pi ^2 q^2\right)}{3 S^2 \omega ^2-\pi  S+3 \pi ^2 q^2},~~
C_{\phi} =\frac{2 S \left(3 S^2 \omega ^2+\pi  S-\pi ^2 q^2\right)}{3 S^2 \omega ^2-\pi  S+\pi ^2 q^2}.\label{rncqcphi}
\end{eqnarray}

It may be observed that both the heat capacities vanish at the extremal value of charge given by the zero of  the temperature $T$ defined in eq.(\ref{rntemp}) as,
\begin{equation}
q_{ex}=\frac{\sqrt{S\left(3 S \omega ^2+\pi  \right)}}{\pi }.\label{rnQex}
\end{equation}

The charges obtained from the divergence of $C_{q}$ and $C_{\phi}$ defined in the euation(\ref{rncqcphi}) are respectively,
\begin{eqnarray}
q_1(S)=\frac{\sqrt{S\left(\pi  -3 S \omega ^2\right)}}{\sqrt{3} \pi },~~
q_2(S)=\frac{\sqrt{S\left(\pi  -3 S \omega ^2\right)}}{\pi }.\label{rnQcqcphi}
\end{eqnarray}
 
 As we have mentioned earlier the thermodynamics and phase structure of black holes are ensemble dependent. Hence we first describe in brief the thermodynamic behaviour for such charged RN-AdS black holes for both the canonical and the grand canonical ensembles. In the canonical ensemble the mass $M$ or the internal energy is allowed to fluctuate while the electric charge is held at a fixed value. Hence the black hole the is in thermal equillibrium with a heat reservoir at the Hawking temperature $T$. The control parameters for this ensemble are $\left(T, q\right)$ and the thermodynamic potential for this ensemble is the Helmholtz potential $F$ that is given as, 
 \begin{equation}
F =m-TS=\frac{3 \pi ^2 q^2+\pi  S-S^2 \omega ^2}{4 \pi ^{3/2} \sqrt{S}}.\label{rnhelmo}
\end{equation}

For the grand canonical ensemble the charge is also allowed to fluctuate along with the mass $M$ or the internal energy.  Hence for this ensemble the black hole is in both  thermal as well as electrical equilibrium with the reservoir held at a constant temperature $T$ and a constant electric potential $\phi$.The control parameters for this ensemble are $\left(T, \phi\right)$ and the thermodynamic potential  for this ensemble is the Gibbs potential $G$. For the charged AdS braneworld black hole this assumes the form,
\begin{equation}
G=m-TS-q \phi =\frac{-\pi ^2 q^2+\pi  S-S^2 \omega ^2}{4 \pi ^{3/2} \sqrt{S}}=\frac{-\pi \phi^2 S+\pi  S-S^2 \omega ^2}{4 \pi ^{3/2} \sqrt{S}}.\label{rnGibbs}
\end{equation}

The charge obtained from the zeroes of the Gibbs potential is given as,
\begin{equation}
q_3(S)=\frac{\sqrt{S \left(\pi -S \omega ^2\right)}}{\pi }.\label{rnQGibbs}
\end{equation}

\begin{figure}[ht!]
\centering
\begin{minipage}[b]{0.45\linewidth}
\includegraphics[width =2.8in,height=1.8in]{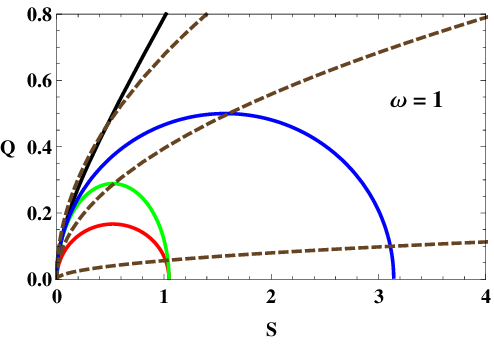}
\end{minipage}%
\begin{minipage}[b]{0.45\linewidth}
\includegraphics[width =2.8in,height=1.8in]{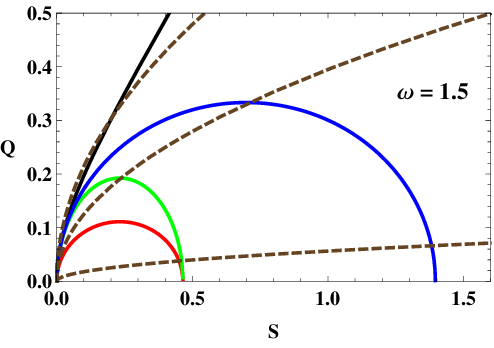}
\end{minipage}\quad
\begin{minipage}[b]{0.45\linewidth}
\includegraphics[width =2.8in,height=1.8in]{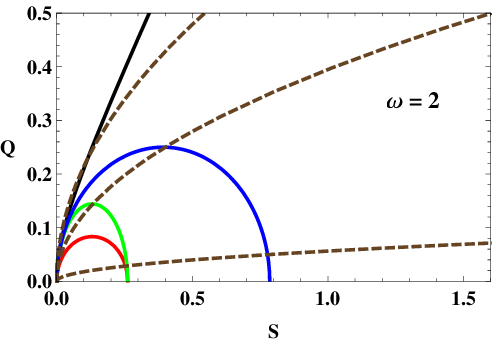}
\end{minipage}%
\begin{minipage}[b]{0.45\linewidth}
\includegraphics[width =2.8in,height=1.8in]{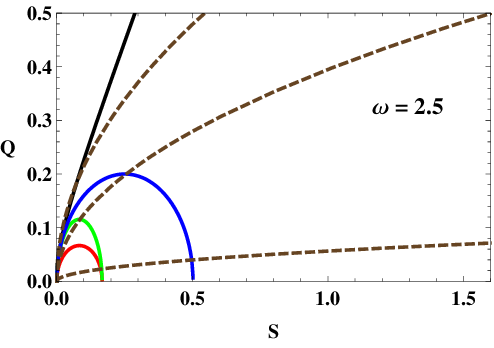}
\end{minipage}\quad
\caption{\label{fig:QSRNads} q vs S Plot for the RN-AdS braneworld black hole at different values of the parameter $\omega=(1,1.5,2,2.5)$. The zeros of the Gibbs free energy are represented by the blue curves. The $C_{\phi}$ and $C_{q}$ spinodal curves are shown in green and red respectively. The extremal curve is shown in the black color and isopotential curves are plotted in dashed brown at $\phi=(0.1,0.7, 1.2)$ from bottom to up.} \end{figure}

In order to study the phase behavior of the system in the two ensembles we first plot the charges $q_1, q_2, q_3$ and the extremal charge $q_{ex}$ with respect to the entropy $S$. The charges $q_1, q_2, q_3$ and $q_{ex}$ are obtained from the divergences of $C_{q}$ , $C_{\phi}$ , Gibbs free energy $\left(G\right) $ and from the zeros of the temperature ie. $\left( T=0 \right)$ respectively. In the fig.(\ref{fig:QSRNads}) we have plotted $q_1, q_2, q_3$ and $q_{ex}$ with respect to entropy $S$ for different values of the braneworld parameter $\omega= \left(1,1.5,2,2.5\right)$. The region left of the extremal curve in black $\left( T=0 \right)$  is the region for a naked singularity and the region on the right side of it is the physical region. The specific heats $C_{\phi}$ and  $C_{q}$ are negative in the inner region bounded by the $C_{q}$  curve in red and $C_{\phi}$ curve in green and outside this region both of them are positive. The Gibbs free energy is positive in the region bounded by blue curve and negative outside it. The brown dotted curves show the isopotential curves for the electric potential $\phi=\left(0.1,0.7,1.2\right)$ from the bottom to the top respectively. 
%%%%%%%%%%%%%%%%%%%%%%%%%%%%%%%%%%%%%%%%%%%%%%%%%%%%%%%%%%%%%%%%%%%%%%%%%%%%%%%%%%%%%%%%
%%%%%%%%%%%%%%%%%%%%%%%%%%%%%%%%%%%%%%%%%%%%%%%%%%%%%%%%%%%%%%%%%%%%%%%%%%%%%%%%%%%%%%%%
In the region bounded by $C_{\phi}$ and $C_{q}$ curves, no stable blackhole solution exists as both of them are negative. For the region bounded between the $C_{\phi}$  curve and the Gibbs curve metastable blackhole solution exist as both $G$ and $C_{\phi}$ remain positive there. The above observations lead to the important conclusion that locally unstable blackhole solutions are also globally unstable as the $C_{\phi}$  curve lies totally inside the region bounded by the Gibbs curve. Outside the Gibbs curve we have both locally and globally stable blackhole solutions. The fig. (\ref{fig:QSRNads}) also depicts that as we increase the value of the parameter $\omega$  the regions bounded by the $C_{\phi}$, $C_{q}$ curves and the Gibbs curve shrinks to a lower area which implies that the regime of instability of the black hole solution decreases with the increase in the value of the parameter $\omega$. It is also possible to analyze the phase coexistence behavior from these curves in both the ensembles. For canonical ensemble we see that a isocharge curve with constant $q$ intersects the $C_{q}$-spinodal curve at two points, which shows that the canonical ensemble displays a liquid-gas like phase behaviour. For $q < q_{c}=\frac{1}{6\omega}$, the RN-AdS braneworld black hole undergoes a first order transition between a small black hole and a large black hole phase charcterized by the horizon radius. At the critical value of the charge $q_{c}=\frac{1}{6\omega}$ corresponding to the maxima of $C_q$ curve, the phase coexistence phenomena ceases to exist and the isocharge curve corresponding to $q_{c}$ is tangential to the $C_q$ curve separating the small black hole branch from the large black hole branch.
%%%%%%%%%%%%%%%%%%%%%%%%%%%%%%%%%%%%%%%%%%%%%%%%%%%%%%%%%%%%%%%%%%%%%%%%%%%%%%%%%%%%%%%%
\begin{figure}[ht!]
\centering
\begin{minipage}[b]{0.4\linewidth}
\includegraphics[width=2.8in,height=1.8in]{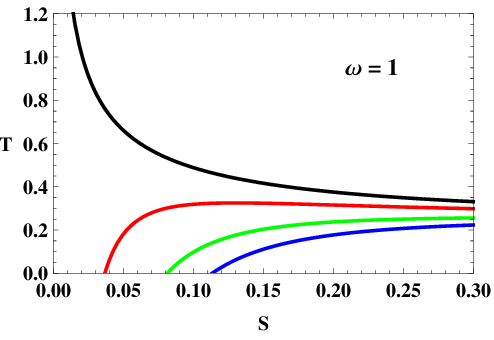}
\caption{$\omega=1$, Black curve q=0, Red curve q=0.11, Green curve q=0.167, Blue curve q=0.2}
\end{minipage}%
\hspace{0.5cm}
\begin{minipage}[b]{0.4\linewidth}
\includegraphics[width=2.8in,height=1.8in]{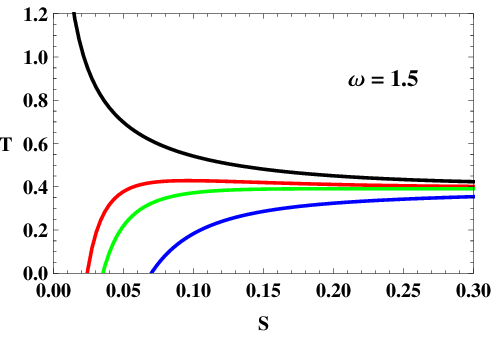}
\caption{$\omega=1.5$, Black curve q=0, Red curve q=0.09, Green curve q=0.11, Blue curve q=0.16}
\end{minipage}\quad
\begin{minipage}[b]{0.4\linewidth}
\includegraphics[width=2.8in,height=1.8in]{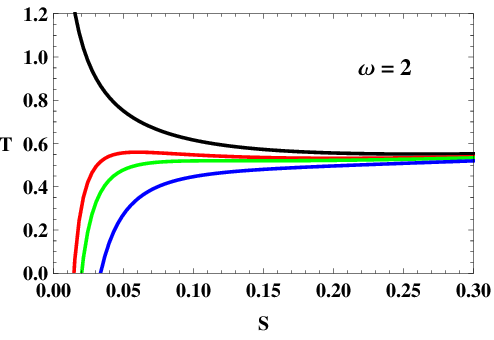}
\caption{$\omega=2$, Black curve q=0, Red curve q=0.07, Green curve q=0.083, Blue curve q=0.11}
\end{minipage}%
\hspace{0.5cm}
\begin{minipage}[b]{0.4\linewidth}
\includegraphics[width=2.8in,height=1.8in]{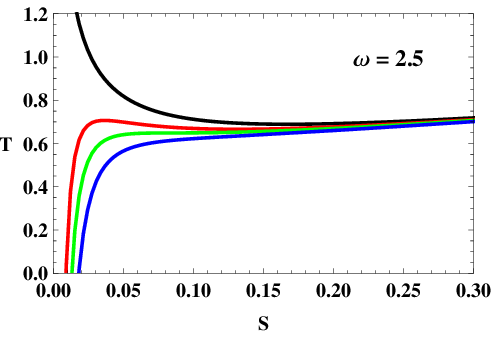}
\caption{$\omega=2.5$, Black curve q=0, Red curve q=0.055, Green curve q=0.067, Blue curve q=0.08}
\end{minipage}%
\end{figure}
In the figures $(2,3,4,5)$  we have plotted the temperature $\left(T\right)$ with respect to the entropy $S$ at fixed values of $q$ $\left(isocharge~curves\right)$ for different values of the parameter $\omega= \left(1,1.5,2,2.5\right)$ respectively. The curve shown in black corresponds to $q=0$ which describes the AdS-Schwarzschild blackhole and exhibits a minima below which there are no stable black hole solutions and above which the stable and the unstable branches coexist at all temperatures. This is known as the 'Davies phase behavior' and the temperature at which this occurs is termed as the Davies temperature $\left(T_{d}\right)$. The effect of adding a finite but fixed charge to the system makes it possible to have a low temperature blackhole branch. At $q_{c}=\frac{1}{6\omega}$ the $T$ vs $S$ curves shows an inflection with critical values of the entropy and the  temperature as $S_{c}=\frac{\pi}{6 \omega^{2}}$ and $T_{c}=\frac{\sqrt{2}}{\pi\sqrt{3}} \omega$ respectively. Note that these critical values show a dependence on the braneworld parameter $\omega$ such that a larger region of stability for the RN-AdS braneworld black hole is observed for higher values of $\omega$. It is also possible to express the temperature $T$ as a function of the entropy $S$ and the electric potential $\phi$ as,
\begin{equation}
T=\frac{\pi +3 S \omega ^2-\pi  \phi ^2}{4 \pi ^{3/2} \sqrt{S}}.\label{rnTsphi}
\end{equation}

\begin{figure}[ht!]
\centering
\begin{minipage}[b]{0.45\linewidth}
\includegraphics[width =2.8in,height=1.8in]{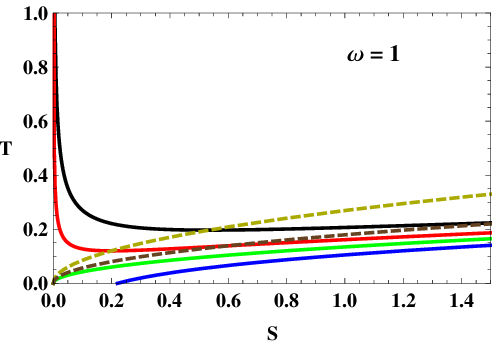}
\end{minipage}%
\begin{minipage}[b]{0.45\linewidth}
\includegraphics[width =2.8in,height=1.8in]{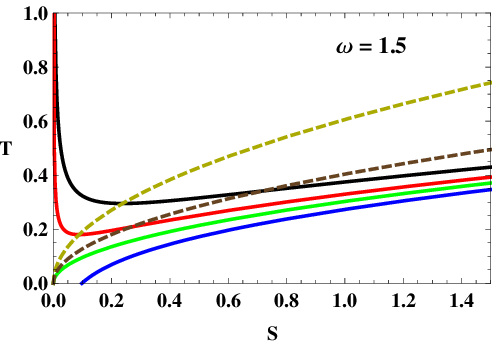}
\end{minipage}\quad
\begin{minipage}[b]{0.45\linewidth}
\includegraphics[width =2.8in,height=1.8in]{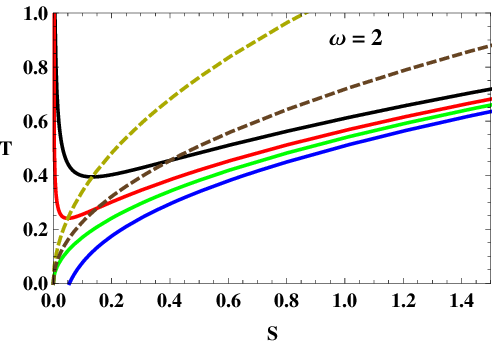}
\end{minipage}%
\begin{minipage}[b]{0.45\linewidth}
\includegraphics[width =2.8in,height=1.8in]{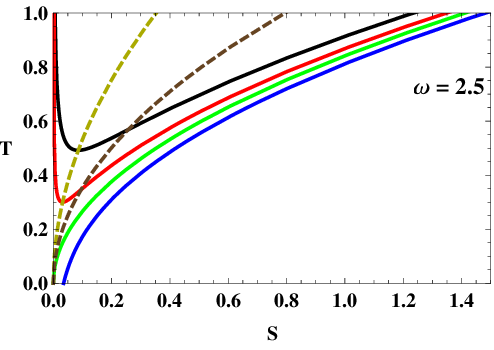}
\end{minipage}%
\caption{\label{fig:TSRNads} Isopotential plots of T vs S for the RN-AdS braneworld black hole at different values of the parameter $\omega$. The curves in black, red ,green and blue stand for values of potential $\phi=(0.7,0.9,1,1.1)$ respectively. The yellow and brown dashed curves are for the divergence of $C_{\phi}$ and $G$ respectively.} \end{figure}

Now in order to investigate the phase behavior in the grand canonical ensemble in the fig.(\ref{fig:TSRNads}) we have plotted the temperature $T$ from the eq.(\ref{rnTsphi}) with respect to the entropy $S$ for fixed values of $\phi$ $\left(isopotential~curves\right)$ and for different values of the braneworld parameter $\omega= \left(1,1.5,2,2.5\right)$. The top two isopotential curves stand for $\phi < 1$ and they show a turning behavior when they cross the $C_{\phi}$ curve indicating the Davies phase transition behavior. Using equations (\ref{rnphi}) and (\ref{rnTsphi}) with  the expression of $Q_2$ from the equation (\ref{rnQcqcphi}), the Davies temperature $T_d$ may be obtained as a function of $\phi$ as follows,
\begin{equation}
T_{d}=\frac{\sqrt{3}}{2}\,{\frac{\omega \sqrt{1-\phi ^2} }{\pi }}.\label{rnTdavies}
\end{equation}

%%%%%%%%%%%%%%%%%%%%%%%%%%%%%%%%%%%%%%%%%%%%%%%%%%%%%%%%%%%%%%%%%%%%%%%%%%%%%%%%%%%%%%%%
%%%%%%%%%%%%%%%%%%%%%%%%%%%%%%%%%%%%%%%%%%%%%%%%%%%%%%%%%%%%%%%%%%%%%%%%%%%%%%%%%%%%%%%%
Eliminating $q$ from equations (\ref{rnphi})  and (\ref{rnQGibbs}) we obtain an expression for the entropy $S$. This expression is substituted in the expression for the temperaturein  (\ref{rnTsphi}) to arrive at the Hawking-Page phase transition temperature $T_h$ that is expressed in terms of the potential $\phi$ as,
\begin{equation}
T_{hp}=\frac{\omega \sqrt{1-\phi ^2} }{\pi }=\frac{2}{\sqrt{3}}T_{d}. \label{rnThawk}
\end{equation}

Thus we observe that for $\phi<1$ the thermal AdS braneworld background is the stable solution for $0<T<T_{d}$, whereas for $T>T_{hp}$ the RN-AdS braneworld black hole is  preferred. In the temperature range $T_{d}<T<T_{hp}$ metastable black hole solutions occur along with the thermal AdS braneworld backgrounds. For $\phi>1$ a single black hole branch exist which is both locally and globally stable for all temperatures and finally at $\phi=1$ this branch has a vanishing entropy at zero temperatures $T=0$. It is to be noted that as we increase $\omega$ the overall stability behavior of the RN-AdS braneworld black hole depending on the Davis temperature $T_{d}$ and the Hawking-page transition temperature $T_{hp}$ remains unchanged.

We now describe the phase behaviour of the RN-AdS braneworld black holes in the framework of thermodynamic geometry. The line element for the thermodynamic geometry for the RN-AdS braneworld black holes may be written as
 \begin{equation}
 \label{rnline}
 dl^2=g_{\mu\nu}dx^{\mu}dx^{\nu}=g_{mm}dm^2+2g_{mq}dmdq+g_{qq}dq^2,
 \end{equation}
 where the indices $\mu, \nu$ in the metric $g_{\mu\nu}$ range over the rescaled variables $m$ and $q$. It may be noted that the Le Chatelier's condition for local thermodynamic stability translates into the constraint of a positive definite line element which in turn requires that the determinant formed by the metric $g_{\mu\nu}$ as well as its principal minors must be  positive definite. Note that for the RN AdS braneworld blackhole, the condition of local thermodynamic stability is satisfied for all the regions lying outside the $C_{\phi}$ curve in the fig.(\ref{fig:QSRNads}). The thermodynamic scalar curvature for the charged RN-AdS braneworld black hole as a function of the entropy $S$ and the electric potential $\phi$ is given as,
\begin{equation}
R=\frac{9 S \omega ^2 \left(-\pi ^2 S^2 \phi ^2+\pi ^2 S^2 \phi ^4-3 \pi  S^3 \omega ^2+4 \pi  S^3 \phi ^2 \omega ^2+3 S^4 \omega ^4\right)}{\left(-\pi  S+\pi  S \phi ^2-3 S^2 \omega ^2\right) \left(-\pi  S+\pi  S \phi ^2+3 S^2 \omega ^2\right)^2}.\label{RNadsR}
\end{equation}

In order to elucidate the behavior and significance of the thermodynamic scalar curvature, we begin by noting that is is possible to express the curvature symbolically  in the following manner
\begin{equation}
R=-9 S \omega ^2 (3 S^2\omega^2+\pi \phi^2 S)\frac{{\cal N}(G)}{{\cal N}(T){\cal D}(C_\phi)^2},\label{RNadsR1}
\end{equation}
here, the symbols ${\cal N}(G),{\cal N}(T)$ and ${\cal D}(C_\phi)$ represent the numerators of the Gibbs potential and the temperature, and the denominator of $C_\phi$  respectively. This clearly shows that the curvature $R$ diverges at extremality and along the $C_\phi$ curve, both of which are at the boundary of the thermodynamically stable  region. Interestingly, it goes to zero along the ``Gibbs curve" of fig.(\ref{fig:QSRNads}) and has opposite sign to the Gibbs free energy everywhere. Now in order to describe the phase behaviour in the grand canonical ensemble, it would be convenient to express the thermodynamic scalar curvature in terms of the potential $\phi$ and the temperature $T$. To achieve this we invert the relation in the equation (\ref{rnTsphi}) to obtain the entropy $S$ as a function of the temperature $ T$  and the electric potential $\phi$ as,
\begin{equation}
S=\frac{\pi  \left(8 \pi ^2 T^2-3 \omega ^2+3 \phi ^2 \omega ^2+4 \pi  \sqrt{4 \pi ^2 T^4-3 T^2 \omega ^2+3 T^2 \phi ^2 \omega ^2}\right)}{9 \omega ^4}.
\end{equation}

Now substituting the expression for $ S$ from the above equation in the expression for the curvature $ R$  in equation (\ref{RNadsR}), one may obtain the curvature $R$ as a function of  $T$ and $\phi$ as,
\begin{eqnarray}
R(T,\phi)& = & \frac{9 \omega ^4 \left(32 \pi ^4 T^4+6 \pi ^2 T^2 \left(-7+8 \phi ^2\right) \omega ^2+9 \left(1-3 \phi ^2+2 \phi ^4\right) \omega ^4\right)}{4 \pi ^2 \left(2 \pi  T^2+Y\right) Z^2} \nonumber \\
&+&\frac{9\omega^4 Y\left(16 \pi ^3 T^2 +3 \pi  \left(-5+6 \phi ^2\right) \omega ^2 \right)}{4 \pi ^2 \left(2 \pi  T^2+Y\right) Z^2},\label{RNadsR2}
\end{eqnarray}
where,
\begin{eqnarray}
Y &=& \sqrt{T^2 \left(4 \pi ^2 T^2+3 \left(-1+\phi ^2\right) \omega ^2\right)}\\
Z &=& 4 \pi ^2 T^2+3 \left(-1+\phi ^2\right) \omega ^2+2 \pi  \sqrt{T^2 \left(4 \pi ^2 T^2+3 \left(-1+\phi ^2\right) \omega ^2\right)},
\end{eqnarray}

\begin{figure}[ht!]
\centering
\begin{minipage}[b]{0.45\linewidth}
\includegraphics[width =2.8in,height=1.8in]{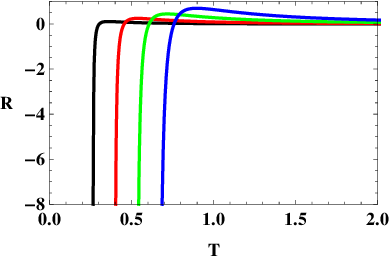}
\end{minipage}%
\begin{minipage}[b]{0.45\linewidth}
\includegraphics[width =2.8in,height=1.8in]{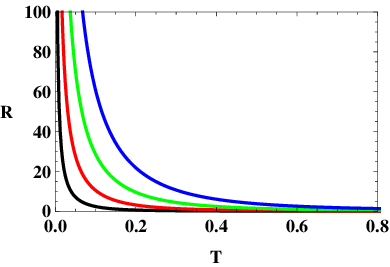}
\end{minipage}%
\caption{\label{fig:RTRNads}Isopotential plots of R vs T for the RN-AdS braneworld black hole at different values of the parameter $\omega$. The curves in black , red, green, blue are for $\omega=(1, 1.5, 2, 2.5)$ at the fixed values of $\phi = 0.3$ (Left) and $\phi = 1.3$ (Right).} \end{figure}

Now  using the expression for the curvature $ R$  as obtained in eq.(\ref{RNadsR2}), we begin with by plotting $R$ with respect to the temperature $T$ in the fig.(\ref{fig:RTRNads}) for various values of the potential $\phi$ and the braneworld parameter $\omega$. These plots describe the phase behavior of the RN-AdS braneworld black hole in the grand canonical ensemble. The left graph in the fig.(\ref{fig:RTRNads}) represents  a generic plot of the scalar curvature $(R)$ vs. the temperature $(T)$ for the values of electric potential $\phi < 1$ and for increasing values of the braneworld parameter $\omega$. It is observed that the curvature $R $ shows a negative divergence at the Davies temperature $T_d$, corresponding to the turning point of the isopotential plots  shown in (\ref{fig:TSRNads}) and it remains negative for the metastable phase of the black hole for  temperatures $T<T_d$. It is further observed that the scalar curvature changes sign to become positive at the Hawking-Page temperature $T_{hp}$. It may be noted that the Davies temperature and the Hawking-page temperature also increase  with the increasing values of the braneworld parameter $\omega$. 
%%%%%%%%%%%%%%%%%%%%%%%%%%%%%%%%%%%%%%%%%%%%%%%%%%%%%%%%%%%%%%%%%%%%%%%%%%%%%%%%%%%%%%%%
%%%%%%%%%%%%%%%%%%%%%%%%%%%%%%%%%%%%%%%%%%%%%%%%%%%%%%%%%%%%%%%%%%%%%%%%%%%%%%%%%%%%%%%%
The right graph in the fig.(\ref{fig:RTRNads}) represents  a generic plot of the thermodynamic  scalar curvature $R$ vs. the temperature $T$ at values of the electric potential $\phi \geq 1$ and for increasing values of the parameter $\omega$. It is observed that the scalar curvature $R$ remains positive at all values of the temperature for $\phi\geq 1$. Thus we conclude that these black holes exist at all temperatures and are locally as well as globally stable. Furthermore, from eq.(\ref{RNadsR2}) it may be seen that  in the limit of large temperature and for all values of the electric potential 
$\phi$, the curvature $R$ asymptotes to zero from the positive side as,
\begin{equation}
R\sim \frac{\omega^4}{T^2}
\end{equation}

We will further elucidate on the issue of the sign of the  scalar curvature $R$ for the RN-AdS braneworld black holes after describing the phase behavior and thermodynamic geometry for the Kerr-AdS braneworld black holes in the next section.

%%%%%%%%%%%%%%%%%%%%%%%%%%%%%%%%%%%%%%%%%%%%%%%%%%%%%%%%%%%%%%%%%%%%%%%%%%%%%%%%%%%%%%%%
\section{Thermodynamic Geometry of the Kerr-AdS Braneworld Black Holes}
In this section, we begin by describing the Kerr-Ads braneworld black holes that are thermodynamically characterized by their mass $M$ and the angular momentum $J$. The Kerr-AdS braneworld black hole are eescribed by the metric described below 
\begin{equation}
ds^{2}=e^{-2A(y)}\left[-\frac{\Delta}{\rho^{2}}(dt-\frac{a\sin^{2}(\theta)}{\Sigma}d\phi)^{2}+\frac{\rho^{2}}{\Delta}dr^{2}+\frac{\rho^{2}}{\Delta_{\theta}}d\theta^{2}+\frac{\Delta_{\theta}\sin^{2}(\theta)}{\rho^{2}}(adt^{2}-\frac{r^{2}+a^{2}}{\Sigma}d\phi)^{2}\right]+dy^{2}
\end{equation}
where,
\begin{eqnarray}
\Delta&=&(r^{2}+a^{2})(1+k^{2}\omega^{2}r^{^{2}})-2\frac{M}{M_{pl}^{2}}r,~~\Delta_{\theta}=1-k^{2}\omega^{2}\frac{a^{2}\cos^{2}(\theta)}{M_{pl}^{4}},\nonumber\\
\rho^{2}&=&r^{2}+\frac{a^{2}\cos^{2}(\theta)}{M_{pl}^{4}},~~\Sigma=1-k^{2}\omega^{2}\frac{a^{2}}{M_{pl}^{4}},\nonumber\\
\end{eqnarray}
here, we have used  $l=\frac{1}{k^{2}\omega^{2}}$. Now using the expression of 
$\Delta$ we determine the event horizon  by solving the equation $\Delta=0$. The solution to the equation leads to  four roots out of which we have only two are physical  corresponding to $r_{+}$ (outer horizon) and $r_{-}$ (inner horizon). Out of the two  $r_{+}=r_{H}$ is used for calculating the entropy of the black hole. 

The entropy $(S)$, angular momentum $(J)$, angular velocity $(\Omega)$ and the mass $(m)$ on the 3-brane for the Kerr-AdS barneworld black holes may be expressed as,
\begin{eqnarray}
S=\pi\frac{(r^{2}+a^{2})}{\Sigma},~ J=\frac{a M}{\Sigma^{2}M_{pl}^{2} },~ \Omega=\frac{\Sigma a}{r^{2}+a^{2}},~ m=\frac{M}{\Sigma M_{pl}^{2}},\label{SJOmdef}
\end{eqnarray}
here we will be working with the rescaled quantities described above. From the expression for $\Delta$ it is possible to determine the temperature $T$, potential $\left(\phi\right)$, the specific heat at constant angular momentum $J$ $\left(C_{J}\right)$ and the specific heat at constant angular velocity $\Omega$ $\left(C_{\Omega}\right)$.  The temperature $T$ may be expressed as a function of the angular momentum $(J)$ and the entropy $(S)$ as,
\begin{equation}
T=\frac{-4 J^2 \pi ^4+S^2 \left(\pi +S \omega ^2\right) \left(\pi +3 S \omega ^2\right)}{4 \pi ^{3/2} \sqrt{S^{\frac{3}{2}}\left(\pi +S \omega ^2\right) \left(4 J^2 \pi ^3+S^2 \left(\pi +S \omega ^2\right)\right)}}.\label{kerrtemp1}
\end{equation}

It is also possible to  express the temperature $(T)$ as a function of the entropy $(S)$ and the angular velocity $(\Omega)$ as,
\begin{equation}
 T=\frac{\pi ^2+\pi  S \left(4 \omega ^2-2 \Omega ^2\right)+3 S^2 \omega ^2 \left(\omega ^2-\Omega ^2\right)}{4 \pi ^{3/2} \sqrt{S \left(\pi +S \omega ^2\right) \left(\pi +S \left(\omega ^2-\Omega ^2\right)\right)}}.\label{kerrtemp2}
\end{equation}

The expression for the angular velocity $(\Omega)$ of black hole may be written as follows,
\begin{equation}
\Omega=\frac{2 J \pi ^{3/2} \left(\pi +S \omega ^2\right)}{\sqrt{S\left(\pi +S \omega ^2\right) \left(4 J^2 \pi ^3+S^2 \left(\pi +S \omega ^2\right)\right)}},\label{kerrOmega}
\end{equation}

The angular velocity $(\Omega)$ as a thermodynamic quantity may be obtained through the differentiation of the Smarr formula. It is to be noted that there is a subtlety in the theromdynamic description of the angular velocity given by eq.(\ref{kerrOmega}). This expression of $\Omega$ is not identical to the angular velocity defined at the horizon  $(\Omega_h)$ which may be observed by considering the limit, $r\rightarrow r_h$ in the expression of $\Omega$ given in eq.(\ref{SJOmdef}). Instead, it turns out to be the difference between the angular velocity at the horizon and the angular velocity defined at the asymptotic boundary of the space-time $(\Omega_\infty)$.  The angular velocity $(\Omega_\infty)$ also turns out to be the angular velocity of the rotating Einstein universe existing at the asymptotic boundary of the Kerr-AdS braneworld black hole. This definition of the angular velocity agrees well with the AdS/CFT picture and describes that the consistent thermodynamics up to the asymptotic boundary is obtained only when $\Omega$ is small, i.e. $\Omega<1$. Inverting the relation of $\Omega$ one obtains an important relation for the angular momentum $J$ as a function of the entropy $S$ and the angular velocity $\Omega$ given as,
\begin{equation}
J=\frac{S^{3/2} \sqrt{-\pi -S \omega ^2} \Omega }{\pi ^{3/2} \sqrt{-4 \pi -4 S \omega ^2+4 S \Omega ^2}}.\label{kerrJ}
\end{equation}

The expressions for heat capacity at constant angular momentum denoted by $C_{J}$ and the heat capacity at constant angular velocity denoted by $C_{\Omega}$ as a function of $J$ and $S$ are as follows,
\begin{eqnarray}
C_{J}&=&\frac{2 S \left(\pi +S \omega ^2\right) \left(4 J^2 \pi ^3+S^2 \left(\pi +S \omega ^2\right)\right) \left(-4 J^2 \pi ^4+S^2 \left(\pi ^2+4 \pi  S \omega ^2+3 S^2 \omega ^4\right)\right)}{S^2 \left(\pi +S \omega ^2\right)^2\left(24 J^2 \pi ^3 \left(\pi +2 S \omega ^2\right)+S^2 \left(\pi +S \omega ^2\right) \left(-\pi +3 S \omega ^2\right)\right)+16 J^4 \pi ^7 \left(3 \pi +4 S \omega ^2\right)},\\
\nonumber\\\nonumber\\
C_{\Omega}&=&-\frac{2 S^3 \left(S \omega ^2+\pi \right)^2 \left(S^2 \left(3 S^2 \omega ^4+4 \pi  S \omega ^2+\pi ^2\right)-4 \pi ^4 J^2\right)}{16 \pi ^8 J^4+8 \pi ^4 J^2 S^2 \left(S \omega ^2+\pi \right)^2+S^4 \left(S \omega ^2+\pi \right)^3 \left(\pi -3 S \omega ^2\right)}
\end{eqnarray}

It can be seen that both the heat capacities vanish at the extremal value of the angular momentum given by the zero of the  temperature T defined in eq.(\ref{kerrtemp1}) as,
\begin{equation}
J_{ex}=\frac{S \sqrt{\left(\pi +S \omega ^2\right)(\pi +3 S \omega )^2} }{2 \pi ^2}.\label{kerrJex}
\end{equation}

We now describe the stability conditions corresponding to these heat capacities. The stability constraint for the canonical ensemble  is obtained from the locus of the divergences of the specific heat $C_J$. Similarly, the stability condition for the grand canonical ensemble is obtained form the divergences of the heat capacity $C_{\Omega}$. The values of the angular momentum $(J)$ obtained from the divergence of $C_{J}$ and $C_{\Omega}$ are given by $J_1$ and $J_2$ respectively as,
\begin{equation}
J_1=\frac{1}{2 \pi ^2}\left(\frac{-3 \pi ^3 S^2-12 \pi ^2 S^3 \omega ^2-15 \pi  S^4 \omega ^4-6 S^5 \omega ^6+2 W}{3 \pi +4 S \omega ^2}\right)^{1/2},\label{kerJ1}
\end{equation}
\begin{equation}
J_2=\frac{S}{2 \pi ^2}\left(2 \sqrt{S \omega ^2 \left(\pi +S \omega ^2\right)^3}- \left(\pi +S \omega ^2\right)^2\right)^{1/2}.\label{kerrJ2}
\end{equation}
where $W$ is given by the expression,
\begin{equation}
W=\sqrt{S^4 \left(\pi +S \omega ^2\right)^3 \left(3 \pi ^3+10 \pi ^2 S \omega ^2+15 \pi  S^2 \omega ^4+9 S^3 \omega ^6\right)}
\end{equation}

We now consider the Kerr-AdS braneworld black hole  in the grand canonical ensemble. For the grand canonical ensemble the internal energy and the angular momentum $(J)$ of the black hole is allowed to fluctuate and the temperature $T $ and the angular velocity $\Omega$ are fixed at equilibrium. A analysis similar to that of the RN-AdS braneworld black holes it is possible to obtain the expressions for the Gibbs free energy and the specific heat at constant angular velocity $C_{\Omega}$ in terms of $S$ and $J$ as,
\begin{equation}
G(J,S) =\frac{4 J^2 \pi ^4+\pi ^2 S^2-S^4 \omega ^4}{4 \pi ^{3/2}\sqrt{S\left(\pi +S \omega ^2\right) \left(4 J^2 \pi ^3+S^2 \left(\pi +S \omega ^2\right)\right)}}.\label{kerrG1}
\end{equation}

The  value of J obtained from the zeros of the Gibbs potential $G$ for the Kerr-AdS braneworld black hole may be given as,
\begin{equation}
J_3=\frac{S \sqrt{-\pi ^2+S^2 \omega ^4}}{2 \pi ^2}
\end{equation}

Substituting the expression for $J$ from the eq.(\ref{kerrJ}) in the  expression for $G$ from eq.(\ref{kerrG1}) one may obtain the Gibbs potential $G$ as a function of the entropy $S$ and the angular velocity $\Omega$ as,
\begin{equation}
 G(S,\Omega)=\frac{\sqrt{S } \left(\pi ^2+S^2 \omega ^2 \left(-\omega ^2+\Omega ^2\right)\right)}{4 \pi ^{3/2} \sqrt{\left(\pi +S \omega ^2\right)\left(\pi +S \left(\omega ^2-\Omega ^2\right)\right)}}.\label{kerrG2}
\end{equation}

Now on eliminating $S$ from the eq.(\ref{kerrtemp2}) and eq.(\ref{kerrG2}), one obtains the Hawking-Page temperature as a function of $\Omega$ as,
\begin{equation}
T_{hp}=\frac{1}{2\pi}\left(2 \omega ^2-\Omega ^2+2 \omega  \sqrt{\omega ^2-\Omega ^2}\right)^{1/2}
\end{equation}
%%%%%%%%%%%%%%%%%%%%%%%%%%%%%%%%%%%%%%%%%%%%%%%%%%%%%%%%%%%%%%%%%%%%%%%%%%%%%%%%%%%%%%%%
\begin{figure}[ht!]
\centering
\begin{minipage}[b]{0.45\linewidth}
\includegraphics[width =2.8in,height=1.8in]{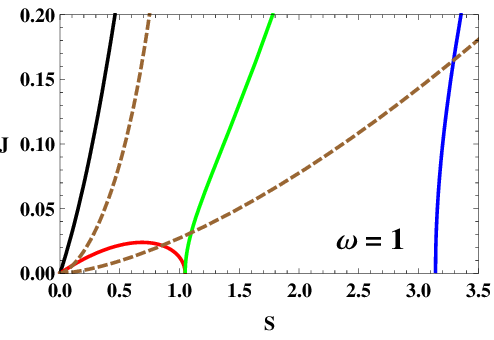}
\end{minipage}%
\begin{minipage}[b]{0.45\linewidth}
\includegraphics[width =2.8in,height=1.8in]{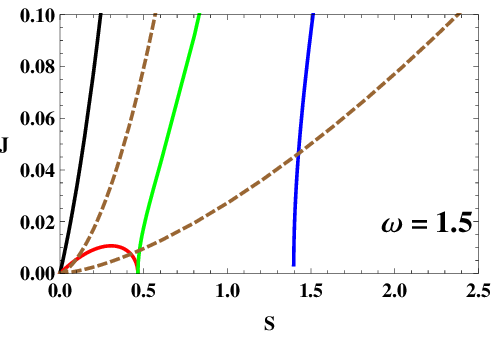}
\end{minipage}\quad
\begin{minipage}[b]{0.45\linewidth}
\includegraphics[width =2.8in,height=1.8in]{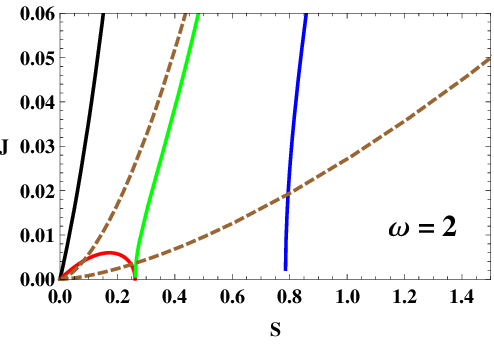}
\end{minipage}%
\begin{minipage}[b]{0.45\linewidth}
\includegraphics[width =2.8in,height=1.8in]{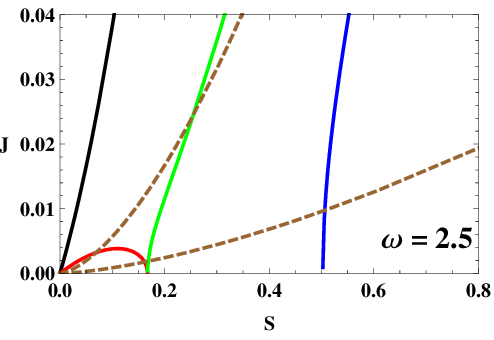}
\end{minipage}%
\caption{\label{fig:JSkerrads} J vs S plots for theKerr-AdS braneworld black hole at different values of the parameter $\omega=(1,1.5,2,2.5)$. The black curve shows the extremal curve with the red and the green curves representing the spinodal curves for $C_{j}$ and $C_{\Omega}$ respectively. The dashed brown curves are isopotentials at $\Omega=0.3$ (lower) and $\Omega=1.9$ (upper).} \end{figure} 
%%%%%%%%%%%%%%%%%%%%%%%%%%%%%%%%%%%%%%%%%%%%%%%%%%%%%%%%%%%%%%%%%%%%%%%%%%%%%%%%%%%%%%%%
Similar to case of the RN-AdS braneworld black holes, we first plot the expressions for the angular momenta $J_1, J_2, J_3$ and $J_{ex}$ with respect to the entropy $S$. It is to be noted that the angular momenta  $J_1, J_2, J_3$ and $J_{ex}$ are obtained from the divergences of $C_{J}$ , $C_{\Omega}$ , Gibbs free energy $\left(G\right) $ and from the zeros of T ie. $\left( T=0 \right)$ respectively. In the figure (\ref{fig:JSkerrads}) we have plotted $J_1, J_2, J_3$ and $J_{ex}$ with respect to entropy $S$ for different values of the parameter $\omega= \left(1,1.5,2,2.5\right)$. The region left to the extremal curves in black $\left( T=0 \right)$, is the region of naked singularity and the region on the right is the physical region. The blue and green curves in the figure correspond to the grand canonical stability curve or the $C_{\Omega}$ curve, to the right of which the heat capacity is positive. The right most blue curves representing the zeros of the Gibbs free energy $(G=0)$, is the Hawking-Page curve, to the right of which the Gibbs free energy of the black hole becomes negative. The red curves connecting the extremal curve and the $C_{\Omega}$ curve is the canonical stability curve or the $C_J$ curve, below which the heat capacity $C_J$ is negative. The brown dotted curves show the isopotential curves $\Omega=\left(0.3, 1.9\right)$ from  bottom to top respectively.

\begin{figure}[ht!]
\centering
\begin{minipage}[b]{0.45\linewidth}
\includegraphics[width =2.8in,height=1.8in]{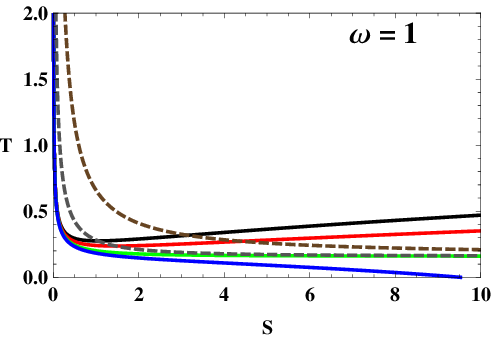}
\end{minipage}%
\begin{minipage}[b]{0.45\linewidth}
\includegraphics[width =2.8in,height=1.8in]{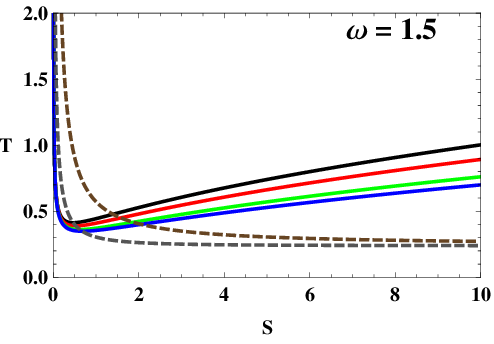}
\end{minipage}\quad
\begin{minipage}[b]{0.45\linewidth}
\includegraphics[width =2.8in,height=1.8in]{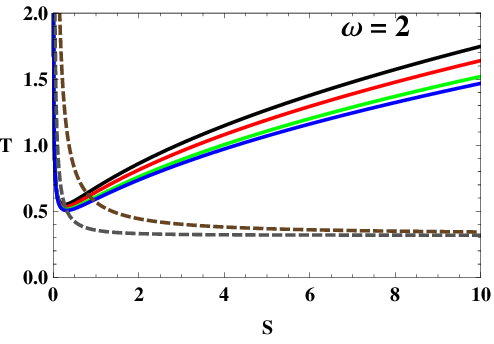}
\end{minipage}%
\begin{minipage}[b]{0.45\linewidth}
\includegraphics[width =2.8in,height=1.8in]{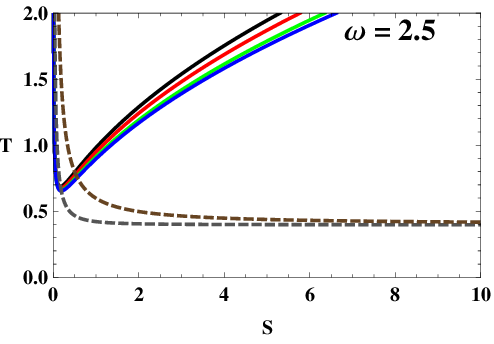}
\end{minipage}%
\caption{\label{fig:TSkerrads} T vs S plots for Kerr-Ads braneworld black hole at different values of the parameter $\omega=(1,1.5,2,2.5)$. The curves in black, red, green and blue are for $\Omega=(0,0.7,1,1.1)$ respectively. The dashed brown and grey curves stand for the divergences of $G$ and $C_{\Omega}$ respectively.} \end{figure}

In the region bounded by $C_{\phi}$ and $C_{q}$ curves, no stable blackhole solution exists as both of them are negative. For the region bounded between the $C_{\phi}$  curve and the Gibbs curve metastable blackhole solutions exist as both $G$ and $C_{\phi}$ remain positive . The above facts lead to the important conclusion that locally unstable blackhole solutions are also globally unstable as the $C_{\phi}$  curve lies completely inside the region bounded by the Gibbs curve. Outside the Gibbs curve we have both locally and globally stable blackhole solutions. The figure (\ref{fig:QSRNads}) also describes that as we increase the value of the braneworld parameter $\omega$  the regions bounded by the $C_{\phi}$, $C_{q}$  curves and the Gibbs curve shrinks to a lower area which implies that the regime of instability of blackhole solutions decrease with the increase in the value of the braneworld parameter $\omega$. For the canonical ensemble we observe that a curve with constant $J$ intersects the $C_{J}$ curve at two points, which shows that the canonical ensemble displays a liquid-gas like phase behaviour similar to the case of the RNAdS braneworld black holes. For $J < J_{c}=\frac{0.33}{\omega^2}$, the Kerr-AdS braneworld black hole undergoes a first order transition between its small black hole and a large black hole phase. At the critical value of charge $J_{c}=\frac{0.33}{\omega^2}$ corresponding to the maxima of $C_J$-spinodal curve, the phase coexistence phenomena ceases to exist and the isocharge curve corresponding to $J_{c}$ becomes tangential to the $C_J$- curve separating the small black hole branch from the large black hole branch (See fig.(\ref{fig:TSkerrads}) ).
%%%%%%%%%%%%%%%%%%%%%%%%%%%%%%%%%%%%%%%%%%%%%%%%%%%%%%%%%%%%%%%%%%%%%%%%%%%%%%%%%%%%%%%%
%%%%%%%%%%%%%%%%%%%%%%%%%%%%%%%%%%%%%%%%%%%%%%%%%%%%%%%%%%%%%%%%%%%%%%%%%%%%%%%%%%%%%%%%
Furthermore, in order to investigate the phase behavior in the grand canonical ensemble in the fig.(\ref{fig:TSRNads}) we begin by plotting the temperature $T$ from the eq.(\ref{kerrtemp2}) with respect to the entropy $S$ for fixed values of $\Omega$ $\left(isopotential~curves\right)$ and for different values of the parameter $\omega= \left(1,1.5,2,2.5\right)$. The dashed grey and brown curves in fig.(\ref{fig:TSRNads}) represent the $C_{\Omega}$-spinodal curves and the zeroes of the Gibbs free energy respectively. The top two isopotential curves shown in black and red colors, stand for $\Omega < 1$ and they show a turning behavior when they cross the $C_{\Omega}$-spinodal curve indicating the Davies phase transition behavior. Using equations (\ref{kerrtemp2}) and (\ref{kerrOmega}) with  the expression of $J_2$ from the eq.(\ref{kerrJ2}), the Davies temperature $T_d$ may be obtained as follows,
\begin{equation}
T_{d}=\frac{2 S^4 \omega ^4+3 \pi  S^3 \omega ^2+\pi ^2 S^2-\sqrt{S^5 \omega ^2 \left(S \omega ^2+\pi \right)^3}}{2 \pi  S^2 \left(S \omega ^2+\pi \right) \left(2 \sqrt{S \omega ^2 \left(S \omega ^2+\pi \right)}-S^2 \omega ^2\right)^{1/2}}.\label{kerTdavies}
\end{equation}

Thus we observe from the graphs in the fig.(\ref{fig:TSRNads}), that for $(\omega=1,\Omega<1)$ the rotating thermal AdS brane background exists for $0<T<T_{d}$, whereas in the temperature range $T_{d}<T<T_{hp}$  the black hole is metastable with the rotating thermal AdS brane background being preferred as it has a lower free energy. For $T>T_{hp}$ the Kerr-AdS braneworld black hole is globally stable and is preferred over the rotating thermal AdS brane background. It is also to be noted that the $\Omega=1$ isopotential curve asymptotes to the $C_{\Omega}$ curve and the Gibbs curve  as the entropy $(S)$ tends to infinity. This clearly shows that for $(\omega=1,\Omega>1)$  the Kerr-AdS braneworld black hole is both locally as well as globally unstable at all temperatures. The remarkable observation from the fig.(\ref{fig:TSRNads}) is that we increase the value of $\omega$ beyond unity,  the Kerr-AdS braneworld black hole becomes both locally as well as globally stable at all temperatures even for values of the angular velocity $(\Omega >1)$ .

We now proceed to discuss the thermodynamic geometry of the Kerr-AdS braneworld black holes by first writing down the thermodyanmic metric  in terms of the  thermodynamic variables which are the angular momentum $(J)$ and the mass $(m)$ both of which are allowed to fluctuate. The line element for the geometry may be written as
\begin{equation}
 dl^2=g_{\mu\nu}dx^{\mu}dx^{\nu}=g_{mm}dm^2+2g_{mJ}dmdJ+g_{JJ}dJ^2,\label{kerrline}
\end{equation}

The scalar curvature corresponding to the thermodynamic  metric is a lengthy expression which may be expressed  as
\begin{eqnarray}
 R&=&\left(\pi  S \left(48 J^4 \pi ^8+S^4 \left(\pi -9 S \omega ^2\right) \left(\pi +S \omega ^2\right)^3+8 J^2 \pi ^4 S^2 \left(\pi +S \omega ^2\right) \left(2 \pi +S \omega ^2\right)\right) \right)\nonumber \\
&&\left(\left(S^2 \left(\pi +S \omega ^2\right)^2+4 J^2 \pi ^3 \left(\pi +2 S \omega ^2\right)\right) \left(4 J^2 \pi ^4 \left(\pi +2 S \omega ^2\right)+S^2 \left(\pi +S \omega ^2\right)^2 \left(\pi +6 S \omega ^2\right)\right)\right)\nonumber \\
&&\left(16 J^4 \pi ^8+8 J^2 \pi ^4 S^2 \left(\pi +S \omega ^2\right)^2+S^4 \left(\pi -3 S \omega ^2\right) \left(\pi +S \omega ^2\right)^3\right)^{-2}\nonumber\\
&&\left(4 J^2 \pi ^4-S^2 \left(\pi +S \omega ^2\right) \left(\pi +3 S \omega ^2\right)\right)^{-1}.\label{keradsR}
\end{eqnarray}

\begin{figure}[ht!]
\centering
\begin{minipage}[b]{0.45\linewidth}
\includegraphics[width =2.8in,height=1.8in]{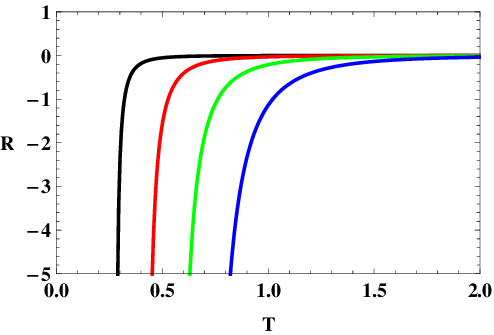}
\end{minipage}%
\begin{minipage}[b]{0.45\linewidth}
\includegraphics[width =2.8in,height=1.8in]{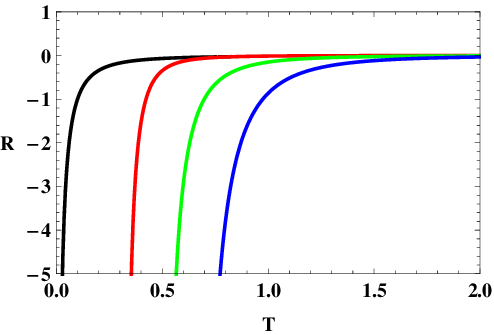}
\end{minipage}%
\caption{\label{fig:RTkerrads} Isopotential plots of R vs T for the Kerr-AdS braneworld black hole at different values of the parameter $\omega$. The curves in black , red, green, blue are for $\omega=(1,1.5,2,2.5)$ at the fixed values of the potential $\Omega=0.03$ (Left) and  $\Omega=1.3$ (Right).} \end{figure}

Once again in order to elucidate the behavior of the scalar curvature, we begin by noting that it is possible to express the curvature symbolically in the following manner
\begin{equation}
R=\frac{{\cal P}_{kerr}}{{\cal N}(T){\cal D}(C_\Omega)^2},\label{keradsR1}
\end{equation}
here, ${\cal P}_{kerr}$ represents a lengthy expression in terms of the entropy $(S)$, angular momentum $(J)$ and the braneworld parameter $\omega$. The symbols ${\cal N}(T)$ and ${\cal D}(C_\Omega)$ represent the numerator of the temperature $(T)$ from eq.(\ref{kerrtemp1}), and the denominator of $C_\Omega$ respectively. From eq.(\ref{keradsR1}) it may be seen that similar to the case of RN-AdS braneworld black hole the scalar curvature encodes the divergences in the grand canonical ensemble for the Kerr-AdS braneworld black hole (See fig.(\ref{fig:RTkerrads}) ).
%%%%%%%%%%%%%%%%%%%%%%%%%%%%%%%%%%%%%%%%%%%%%%%%%%%%%%%%%%%%%%%%%%%%%%%%%%%%%%%%%%%%%%%%
%%%%%%%%%%%%%%%%%%%%%%%%%%%%%%%%%%%%%%%%%%%%%%%%%%%%%%%%%%%%%%%%%%%%%%%%%%%%%%%%%%%%%%%%
For the RN-AdS braneworld black hole it is seen that the curvature describes a crossing behavior for $\phi<1$ when plotted with respect to the temperature i.e., it changes sign from -ve to +ve at the Hawking-page temperature $T_{hp}$ and asymptotes to zero from the positive side as the temperature $T$ goes  to infinity. This crossing behavior of the curvature is not observed for the Kerr-AdS braneworld black holes and thus it does
not describe the signature for the Hawking-Page phase transition. In fact the curvature always asymptotes to zero from the negative side as temperature $(T)$ tends to infinity. This empahsizes an important distinction between the  scalar curvatures for the two different black holes discussed here. It is possible further to obtain the curvature $R$ as a function of $T$ and $S$ using  eq.(\ref{kerrtemp2}), eq.(\ref{kerrJ}) and eq.(\ref{keradsR}) for the Kerr-AdS braneworld black holes. Using the expression for the curvature $R$ in terms of  $T$ and $S$ we may obtain the asymptotic behavior of $R$ in the limit of large temperature as,
\begin{equation}
R\sim-\frac{\omega^4}{T^4}
\end{equation}

At this point it is worth to mention that the exact significance of the sign of the state space scalar curvature $(R)$ is still an unsettled issue in the context of thermodynamic geometry of the black hole systems. Thus in this section we have discussed the thermodynamic geometry of the Kerr-AdS braneworld black holes and also we have prompted out the major difference in the behavior of the state space scalar curvatures for the two black hole systems taken here.

%%%%%%%%%%%%%%%%%%%%%%%%%%%%%%%%%%%%%%%%%%%%%%%%%%%%%%%%%%%%%%%%%%%%%%%%%%%%%%%%%%%%%%%%
\section{Summary and Discussions}
In this article we have described the thermodynamics and phase structures for both the charged RN-AdS and the rotating Kerr-AdS braneworld black holes in a generalized Randall-Sundrum construction in the framework of thermodynamic geometry. The generalized RS model envisages a non zero cosmological constant on the three brane which is embedded in a slice of $AdS_5$ with the usual orbifold compactification of the extra fifth dimension on $\frac {S^1}{Z_2}$. This naturally allows for both dS and AdS braneworld black hole solutions on the three brane. We implement a careful analysis of the thermodynamics, stability and phase structure in both the canonical and the grand canonical ensemble for the RN AdS and the Kerr-AdS braneworld black holes. This involved the determination of the specific heats as functions of the thermodynamic variables and also the corresponding  entropy and the free energy relevant to the respective ensembles. A detailed graphical analysis for the behaviour of the specific heat and the free energy shows the dependence of the thermodynamic behaviour and stability on the braneworld parameters. We then compute the thermodynamic metric and the scalar curvature for both the RN-AdS and the Kerr-AdS braneworld black holes respectively in the grand canonical ensemble. The thermodynamic scalar curvatures are analyzed graphically as functions of the temperature to study the phase structures of the AdS braneworld black holes. Our graphical analysis describes both the Hawking Page and the Davies turning point behaviour for the AdS braneworld black holes in the grand canonical ensemble. Interestingly the curvature exhibits an interesting dependence on the braneworld parameter. We observe that in general we have larger regions of stability for these braneworld black holes for higher values of the brane world parameter which is related to the cosmological constant on the three brane.

The thermodynamics stability and the phase structures for a charged rotating Kerr-Newman AdS braneworld black hole in a generalized RS construction is an interesting issue for future investigations. These black holes with a larger number of thermodynamic parameters maybe described in mixed canonical grand canonical ensembles. This may lead to extremely interesting phase structures involving both first order and second order transitions observed for such AdS black holes in conventional scenarios. The study of the thermodynamic geometry and the curvature for these black holes in such mixed ensembles may lead to interesting new insights into the phase structures of such AdS braneworld black holes. These issues are left for a future study.

\begin{appendices}
\section{A note on the derivation of the bareworld black hole solutions.}
In this article we have considered the 4- dimensional world $({\cal M})$ to be described by a domain wall (3-brane) embedded in a 5- dimensional spacetime $({\cal V},g_{\mu\nu})$ such that the Greek indices, $\{\mu,\nu,\cdots\}=\{0,1,2,3,4\}$ run over the full spacetime. If one denotes the vector normal to the 3-brane $({\cal M})$ by $n^\alpha$, then the induced metric on the 3-brane may be given by, $\tilde{g}_{\mu\nu}=g_{\mu\nu}-n_{\mu}n_{\nu}$. We begin with the Gauss and Codacci equations which may be given as 
\begin{eqnarray}
{}^{(4)}{\cal R}^{\alpha}_{\beta\gamma\delta}&=&{}^{(5)}{\cal R}^{\mu}_{\nu\rho\sigma}\tilde{g}_{\mu}^{\alpha}\tilde{g}_{\beta}^{\nu}\tilde{g}_{\gamma}^{\rho}\tilde{g}_{\delta}^{\sigma}+{\cal K}_{\gamma}^{\alpha}{\cal K}_{\beta\delta}-{\cal K}_{\delta}^{\alpha}{\cal K}_{\beta\gamma},\label{Gausseq}\\
{}^{(5)}{\cal R}_{\rho\sigma}n^{\sigma}\tilde{g}_{\mu}^{\rho}&=&{\cal D}_{\nu}{\cal K}_{\mu}^{\nu}-{\cal D}_{\mu}{\cal K},\label{Codaccieq}
\end{eqnarray}
here, ${\cal K}={\cal K}_{\mu}^{\mu}$ denotes the extrinsic curvature of ${\cal M}$ such that, ${\cal K}_{\mu\nu}=\tilde{g}_{\mu}^{\alpha}\tilde{g}_{\nu}^{\beta}\nabla_{\alpha}n_{\beta}$ and ${\cal D}_{\mu}$ is the covariant differentiation respect to $\tilde{g}_{\mu\nu}$. Now, contracting the Gauss equation in 
eq.(\ref{Gausseq}) for $\alpha$ and $\gamma$ one may obtain
\begin{equation}
{}^{(4)}{\cal R}_{\mu\nu} = {}^{(5)}{\cal R}_{\rho\sigma} \tilde{g}_\mu^{~\rho}\tilde{g}_\nu^{~\sigma} - {}^{(5)}{\cal R}^\alpha_{~\beta\gamma\delta}n_\alpha \tilde{g}_\mu^{~\beta} n^\gamma \tilde{g}_\nu^{~\delta}  + {\cal K}{\cal K}_{\mu\nu} -{\cal K}^{~\alpha}_{\mu}{\cal K}_{\nu\alpha},
\label{Ricci}
\end{equation}
which further gives us
\begin{eqnarray}
{}^{(4)}G_{\mu\nu}&=&\left[{}^{(5)}{\cal R}_{\rho\sigma} -\frac{1}{2} g_{\rho\sigma}{}^{(5)}{\cal R}\right] \tilde{g}_\mu^{~\rho} \tilde{g}_\nu^{~\sigma}+{}^{(5)}{\cal R}_{\rho\sigma}n^\rho n^\sigma \tilde{g}_{\mu\nu}  +{\cal K}{\cal K}_{\mu\nu} -{\cal K}^{~\rho}_{\mu}{\cal K}_{\nu\rho}\nonumber\\
&-&\frac{1}{2}\tilde{g}_{\mu\nu}  ({\cal K}^2-{\cal K}^{\alpha\beta}K_{\alpha\beta}) -  {}^{(5)}{\cal R}^\alpha_{~\beta\rho\sigma}n_\alpha n^\rho 
\tilde{g}_\mu^{~\beta} \tilde{g}_\nu^{~\sigma}
\label{4dEinsteineq}
\end{eqnarray}

Moreover, we have the 5-dimensional Einstein equations,
\begin{equation}
{}^{(5)}{\cal R}_{\alpha\beta} -\frac{1}{2}g_{\alpha\beta}{}^{(5)}{\cal R} =\kappa_5^2\,{\cal T}_{\alpha\beta}
\label{5dEinstein}
\end{equation}
where, ${\cal T}_{\mu\nu}$ is the 5-dimensional energy-momentum tensor. The 5-dimensional Einstein equations together with the following decomposition of the Riemann tensor into the Weyl curvature, the Ricci tensor and the scalar curvature
\begin{eqnarray}
{}^{(5)}{\cal R}_{\mu\alpha\nu\beta}=\frac{2}{3}( g_{\mu [\nu}{}^{(5)}R_{\beta]\alpha} -g_{\alpha [ \nu}{}^{(5)}{\cal R}_{\beta] \mu})-\frac{1}{6}g_{\mu [\nu}g_{\beta ]\alpha}{}^{(5)}{\cal R}+{}^{(5)}{\cal C}_{\mu\alpha\nu\beta}, 
\end{eqnarray}
results into the following form of the 4-dimensional Einstein equations
\begin{eqnarray}
{}^{(4)}G_{\mu\nu}&=& {2 \kappa_5^2 \over 3}\left({\cal T}_{\rho\sigma}\tilde{g}^{~\rho}_{\mu}  \tilde{g}^{~\sigma}_{\nu}+\left({\cal T}_{\rho\sigma}n^\rho n^\sigma-{1 \over 4}{\cal T}^\rho_{~\rho}\right) \tilde{g}_{\mu\nu} \right)+ {\cal K} {\cal K}_{\mu\nu}\nonumber\\
&-&{\cal K}^{~\sigma}_{\mu}{\cal K}_{\nu\sigma} -{1 \over 2}\tilde{g}_{\mu\nu}\left( {\cal K}^2-{\cal K}^{\alpha\beta} {\cal K}_{\alpha\beta}\right) -  {}^{(5)}{\cal C}^\alpha_{~\beta\rho\sigma}n_\alpha n^\rho\tilde{g}_\mu^{~\beta}\tilde{g}_\nu^{~\sigma}, 
\label{4dEinstein}
\end{eqnarray}

Uptill now, in the analysis presented above, we have not considered any particular symmetry or a particular form of the energy momentum tensor of the 5-dimensional spacetime. Thus in the the brane world scenario, we choose a coordinate $y$ for which the hypersurface $y=0$ coincides with the 3-brane and $n_\mu dx^\mu=dy$ such that, $n^\nu\nabla_\nu n^\mu=0$ for this extra dimension. More explicitly we assume that in the presence of the 3-brane the 5-dimensional metric and the energy-momentum tensor has the following form
\begin{eqnarray}
ds^2=dy^2+\tilde{g}_{\mu\nu}dx^\mu dx^\nu,~~~{\cal T}_{\mu\nu}=-\Lambda g_{\mu\nu}+S_{\mu\nu}\delta (y), \label{eq:bulk}
\end{eqnarray}
where, 
\begin{equation}
S_{\mu\nu}=-\lambda \tilde{g}_{\mu\nu}+\tau_{\mu\nu},~~\tau_{\mu\nu}n^{\nu}=0
\end{equation}
here, $\Lambda$ is the cosmological constant of the 5- dimensional spacetime. Furthermore,  $\lambda$ and $\tau_{\mu\nu}$ correspond to the vacuum energy and the energy-momentum tensor, respectively, for the 3-brane. Following \cite{Shiromizu:1999wj} one may impose a $Z_2$ symmetry on the 5-dimensional spacetime and using the well known Israel’s junction condition near the hypersurface $y=0$, it may be observed that the extrinsic curvature of the 3-brane can be determined terms of the 4-dimensional energy momentum tensor as follows
\begin{equation}
K_{\mu\nu}=-\frac{1}{2}\kappa_5^2 \Bigl(S_{\mu\nu}-\frac{1}{3}q_{\mu\nu}S\Bigr).\label{eq:zsym}
\end{equation}

Substituting eq.~(\ref{eq:zsym}) into eq.~(\ref{4dEinstein}), one may obtain the gravitational equations on the 3-brane as
\begin{eqnarray}
{}^{(4)}G_{\mu\nu}=-\Lambda_4 \tilde{g}_{\mu\nu}+ 8 \pi G_N\tau_{\mu\nu}+\kappa_5^4\,\pi_{\mu\nu} -{}^{(5)}{\cal C}^\alpha_{~\beta\rho\sigma}n_\alpha n^\rho\tilde{g}_\mu^{~\beta}\tilde{g}_\nu^{~\sigma}, \label{eq:effective}
\end{eqnarray}
where,
\begin{eqnarray}
\Lambda_4&=&\frac{1}{2}\kappa_5^2 \left(\Lambda +\frac{1}{6}\kappa_5^2\,\lambda^2\right),\label{Lamda4}\\
G_N&=&{\kappa_5^4\,\lambda\over48 \pi},\label{GNdef}\\
\pi_{\mu\nu}&=& -\frac{1}{4} \tau_{\mu\alpha}\tau_\nu^{~\alpha}+\frac{1}{12}\tau\tau_{\mu\nu}+\frac{1}{8}q_{\mu\nu}\tau_{\alpha\beta}\tau^{\alpha\beta}-\frac{1}{24}q_{\mu\nu}\tau^2,\label{pidef}
\end{eqnarray}
here, $\Lambda_4$ is the four dimensional cosmological constant on the 3-brane. Moreover, it may also be seen that the last term in the eq.~(\ref{eq:effective}) is a part of the 5-dimensional Weyl tensor and carries information of the gravitational field outside the brane.
This term is non-vanishing if the 5-dimensional spacetime is not purely Anti-de Sitter and it is constrained by the motion of the matter on the brane \cite{Shiromizu:1999wj}. It may also be observed that a metric of the form given in eq.(\ref{metric2}) satisfies both the equations (\ref{eq:effective}) and (\ref{5dEinstein}) and the metric on the 3-brane can be obtained by solving the 4-dimensional Einstein's equation given by (\ref{eq:effective}). Thus on the 3-brane one may specify a metric corresponding to a charged $AdS$ or a Kerr-AdS black hole.
\end{appendices}

%%%%%%%%%%%%%%%%%%%%%%%%%%%%%%%%%%%%%%%%%%%%%%%%%%%%%%%%%%%%%%%%%%%%%%%%%%%%%%%%%%%%%%%%

%%%%%%%%%%%%%%%%%%%%%%%%%%%%%%%%%%%%%%%%%%%%%%%%%%%%%%%%%%%%%%%%%%%%%%%%%%%%%%%%%%%%%%%%
\end{document}